\documentclass[letter]{aa} 

\usepackage[utf8]{inputenc}

\usepackage{booktabs}
\usepackage{threeparttable}
\usepackage{graphicx}
\graphicspath{{Figures/}}

\usepackage{color}
\usepackage{amsmath}
\usepackage{epstopdf}
\usepackage{url}

\usepackage{xcolor}
\definecolor{green}{HTML}{33CC33}
\definecolor{red}{HTML}{FF3300}
\definecolor{blue}{HTML}{3333FF}

\usepackage[format=plain,labelformat=default, labelsep=endash, labelfont=bf, margin=0pt, tableposition=top, font={footnotesize}]{caption}
\usepackage{subcaption}
\usepackage{times}
\usepackage[english]{babel}
\usepackage{mathrsfs}
\usepackage{xspace}

\newcommand{\ie}{i.e.\@\xspace} 
\newcommand{\eg}{e.g.\@\xspace} 

\newcommand{\Kepler}{\textit{Kepler}\xspace}
\newcommand{\Gaia}{\textit{Gaia}\xspace}

\renewcommand{\eqref}[1]{Equation~\ref{#1}}
\newcommand{\fref}[1]{Figure~\ref{#1}}

\numberwithin{equation}{section}

\makeatletter
\def\maketag@@@#1{\hbox{\m@th\normalfont\normalsize#1}}
\makeatother

\makeatletter
\newcommand\footnoteref[1]{\protected@xdef\@thefnmark{\ref{#1}}\@footnotemark}
\makeatother

\usepackage[varg]{txfonts}
\usepackage{natbib,twoopt}
\usepackage[breaklinks=true,colorlinks=true,citecolor=blue,urlcolor=blue]{hyperref} 
\bibpunct{(}{)}{;}{a}{}{,} 
\makeatletter
\newcommandtwoopt{\citeads}[3][][]{\href{http://adsabs.harvard.edu/abs/#3}%
{\def\hyper@linkstart##1##2{}%
\let\hyper@linkend\@empty\citealp[#1][#2]{#3}}}
\newcommandtwoopt{\citepads}[3][][]{\href{http://adsabs.harvard.edu/abs/#3}%
{\def\hyper@linkstart##1##2{}%
\let\hyper@linkend\@empty\citep[#1][#2]{#3}}}
\newcommandtwoopt{\citetads}[3][][]{\href{http://adsabs.harvard.edu/abs/#3}%
{\def\hyper@linkstart##1##2{}%
\let\hyper@linkend\@empty\citet[#1][#2]{#3}}}
\newcommandtwoopt{\citeyearads}[3][][]%
{\href{http://adsabs.harvard.edu/abs/#3}
{\def\hyper@linkstart##1##2{}%
\let\hyper@linkend\@empty\citeyear[#1][#2]{#3}}}
\makeatother

\defcitealias{joris}{DeR16} 
\defcitealias{thiase}{R14} 
\defcitealias{TGASP}{TGAS} 
\renewcommand{\thesubfigure}{\arabic{subfigure}}

\newcommand{\guy}[1]{{\color{black} #1}}

\title{Using red clump stars to correct the \Gaia DR1 parallaxes}
\author{Guy~R.~Davies\inst{\ref{inst1},\ref{inst2}}\thanks{\email{g.r.davies@bham.ac.uk}}\and 
Mikkel~N.~Lund\inst{\ref{inst2},\ref{inst1}} \and 
Andrea~Miglio\inst{\ref{inst1},\ref{inst2}} \and
Yvonne~Elsworth\inst{\ref{inst1},\ref{inst2}} \and 
James~S.~Kuszlewicz\inst{\ref{inst1},\ref{inst2}} \and 
Thomas~S.H.~North\inst{\ref{inst1},\ref{inst2}} \and 
Ben~Rendle\inst{\ref{inst1},\ref{inst2}} \and 
William~J.~Chaplin\inst{\ref{inst1},\ref{inst2}} \and 
Tha\'ise~S.~Rodrigues\inst{\ref{inst3},\ref{inst4}} \and
Tiago~L.~Campante\inst{\ref{inst1},\ref{inst2}} \and 
L\'eo~ Girardi\inst{\ref{inst3}}\and
Steven~J.~Hale\inst{\ref{inst1},\ref{inst2}} \and 
Oliver~Hall\inst{\ref{inst1},\ref{inst2}} \and 
Caitlin~D.~Jones\inst{\ref{inst1},\ref{inst2}} \and 
Steven~D.~Kawaler\inst{\ref{inst5}} \and
Ian~Roxburgh\inst{\ref{inst6}} \and 
Mathew~Schofield\inst{\ref{inst1},\ref{inst2}}
}

\institute{School of Physics and Astronomy, University of Birmingham, Edgbaston, Birmingham, B15 2TT, UK\label{inst1}
\and Stellar Astrophysics Centre, Department of Physics and Astronomy, Aarhus University, Ny Munkegade 120, DK-8000 Aarhus C, Denmark\label{inst2}
\and Osservatorio Astronomico di Padova -- INAF, Vicolo dell'Osservatorio 5, I-35122 Padova, Italy\label{inst3}
\and Dipartimento di Fisica e Astronomia, Universit\`a di Padova,  Vicolo dell'Osservatorio 2, I-35122 Padova, Italy\label{inst4}
\and Department of Physics and Astronomy, Iowa State University, Ames, IA 50011, USA\label{inst5} 
\and Astronomy Unit, Queen Mary University of London, Mile End Road, London E1 4NS, UK\label{inst6}
}

\date{Received 11. Nov. 2016 / Accepted 10. Jan. 2017}

\abstract{
Recent results have suggested that there is tension between the \Gaia DR1 TGAS distances and the distances obtained using luminosities determined by eclipsing binaries or asteroseismology on red giant stars. We use the $K_s$-band luminosities of red clump stars, identified and characterized by asteroseismology, to make independent distance estimates. Our results suggest that \Gaia TGAS distances contain a systematic error that decreases with increasing distance. We propose a correction to mitigate this offset as a function of parallax that is valid for the \Kepler field and values of parallax that are less than ${\sim} 1.6 \rm \, mas$.  For parallaxes greater than this we find agreement with previously published values.  \guy{We note that the TGAS distances to the red clump stars of the open cluster M67 show a high level of disagreement that is difficult to correct for.}}


\begin{document}
\maketitle
\section{Introduction}\label{sec:intro}
The {\it Gaia} mission promises trigonometric parallaxes for ${\sim}10^{9}$
stars with precisions of tens of micro arcseconds.  The first data
release (DR1) -- the {\it Tycho-Gaia} Astrometric Solutions sample
\citep[][]{2015A&A...574A.115M,TGASP} (hereafter \citetalias{TGASP}) --
is based on only 14 months of data and provides parallax estimates for
some 2 million stars. Initial comparisons have suggested that the TGAS
sample contains a systematic offset of around $-0.25 \, \rm mas$
\citep[][]{2016arXiv160905390S, 2016arXiv161100656J}\guy{, or that 
no correction is required \citep[][]{2016arXiv160904303L, 2016arXiv161107035S}.}  Here
we test the TGAS parallaxes using a sample of stars observed by the
NASA {\Kepler} mission that provide a useful probe of the far end of the
distance scale in the {\it Kepler} field of view.
\par
\begin{figure}[ht]
\includegraphics[width=\columnwidth]{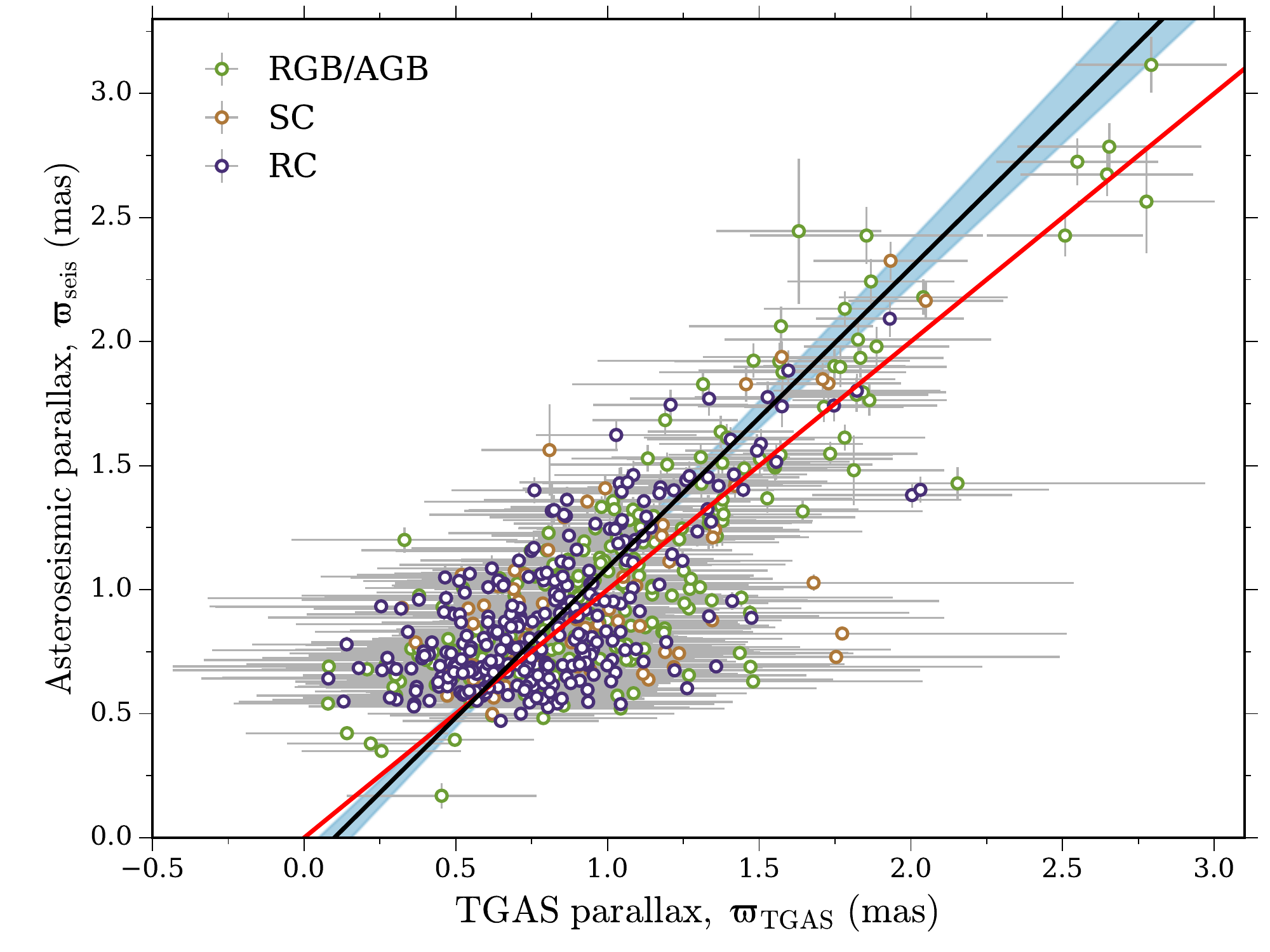}

\caption{Comparison between asteroseismic and TGAS parallaxes. The
  markers indicate the evolutionary state of the star, including the
  red giant branch (RGB or AGB), the red clump (RC), and the secondary clump
  (SC). The red line shows the 1:1 relation. The black line shows
  the linear relation obtained from an orthogonal distance regression
  (ODR), which includes uncertainties on both parallax estimates. The
  best-fitting relation is $\varpi_{\rm seis} = (1.21\pm 0.07)
  \times \varpi_{\rm TGAS} -(0.12\pm 0.08)$. The shaded blue region
  shows the 1-$\sigma$ confidence interval around the best-fitting
  relation.}

\label{fig:par}
\end{figure}

 
\citet[][]{joris} (hereafter \citetalias{joris}) have recently shown
that there is a tension between the astrometric distances of the TGAS
sample and asteroseismically determined distances of red giants from
\citet[][]{thiase} (hereafter \citetalias{thiase}).  A linear fit of
the asteroseismic parallaxes to the TGAS parallaxes returns a slope
that is significantly different from unity, and an intercept that is
significantly different from zero.  We have replicated the
\citetalias{joris} result in \fref{fig:par} using an orthogonal
distance regression \guy{(ODR, \cite[][]{boggs1990orthogonal})}, and find comparable results. It has been
suggested by \citetalias{joris} that the above departures indicate
that either the TGAS parallaxes are biased; or that asteroseismic parallaxes are
compromised by incorrect interstellar extinction corrections and/or 
poorly known bulk metallicities, which would introduce systematics in the
estimated stellar luminosities.
\par 
Here, we have tested the TGAS and asteroseismic distance scales independently by
studying a subset of the stars from \citetalias{joris}, specifically
those stars that have been robustly identified by asteroseismology as
red clump (designated RC and defined to include core helium-burning stars but 
exclude red giant branch (RGB), asymptotic giant branch (AGB), and secondary red clump (SC)
stars). We also added a further six RC stars from the open cluster M67,
which have asteroseismic detections \citep{M67} in data collected by the
re-purposed {\Kepler} Mission, K2. 

We make use of an important property of the RC stellar population -- specifically that 
RC stars have more or less the same luminosity 
\citep[\eg,][]{1970MNRAS.150..111C, 1998ApJ...494L.219P} -- to validate the estimated 
distances. As shown by \citet[][]{2002MNRAS.337..332S} and \citet[][]{RCstars}, 
the $K_s$-band minimizes the intrinsic differences in the luminosites 
of RC stars due to differences in their metallicities. Residual changes due to variations 
in stellar masses, ages, and evolution along the He-burning phase remain at the level 
of ${\lesssim} 0.2$ mag. The main limitation in the use of RC stars as 
distance indicators stems from the difficulty in identifying them among 
the wider red giant population (with the exception of ensembles of 
stars located at similar distances, e.g. in clusters, the 
Galactic bulge, or nearby galaxies). This limitation has now been 
overcome thanks to asteroseismic constraints that can unambiguously discern 
RC from SC and RGB or AGB stars \citep{2011Natur.471..608B}.  Hence RC stars 
can be used to test \Gaia distances. 
\section{The asteroseismic constraints}
\label{sec::seismic}
Asteroseismology provides two key sets of constraints which may be used as independent ways of inferring distances: 
\begin{itemize}
\item[$\circ$] Radii of red giant stars can be estimated from the average asteroseismic parameters that characterize their acoustic oscillation spectra: the so-called average large frequency separation, and the frequency corresponding to the maximum observed oscillation power.
Red giants, which show these solar-like oscillations, may therefore be used as accurate distance indicators just as in the case of eclipsing binaries: the distance to each red giant may be estimated from the absolute luminosity, which is obtained from the asteroseismically determined radius (for example from asteroseismic scaling relations) and $T_{\rm eff}$ \citep[e.g.,][]{2013MNRAS.429..423M}.
\item[$\circ$] Thanks to the frequencies of dipolar gravito-acoustic modes  we can discern pristine RC stars among the zoo of red-giant stars \citep[][]{2011Natur.471..608B}. We note
that these inferences are independent of the constraints used to determine radii (and hence distances). Once RC stars are identified, their distances may be determined given their intrinsic luminosity.
\end{itemize}
RC stars in the solar neighborhood are expected to have similar intrinsic
$K_s$-band luminosities (to within ${\sim} 0.2$ mag), so we anticipate a RC 
coherent, extended feature in a diagram
showing distance and apparent luminosity.  This feature would show the
expected distance dependence subject to some scatter or offset from
the effects of reddening (photometric band dependent), stellar
multiplicity, and the intrinsic scatter of RC luminosities.  
\par
We adopt a set of asteroseismic distances calculated using a model-independent method,
i.e., using the asteroseismic scaling relations \citep[e.g.,][]{2013MNRAS.429..423M}; 
they are tabulated as the ``distances from direct method'' in the online data 
provided by \citetalias{thiase}.  \guy{This method has no knowledge of the evolutionary 
state of the star and being model-independent does not force the estimated luminosities 
to some assumed RC value.}  This ensures that we have an
independent estimate of luminosity for each star to use as a test of
the assumption that the RC has a low scatter in luminosity.  
 
\par 
We selected our RC sample using an asteroseismic classification
\citep[][]{Elsworth2016} that exploits the diagnostic
properties of dipole modes of mixed character, allowing one to
discriminate between hydrogen-shell and helium core-burning red giant
stars \citep[][]{2011Natur.471..608B}. This classification has been shown to be
robust in the identification of RC stars and is capable of separating
out secondary clump stars and with the addition of temperature from
\citet[][]{2014ApJS..215...19P}, the horizontal branch stars.
\section{Results}
\label{sec::results}
\begin{figure*}[ht]
\includegraphics[width=\textwidth]{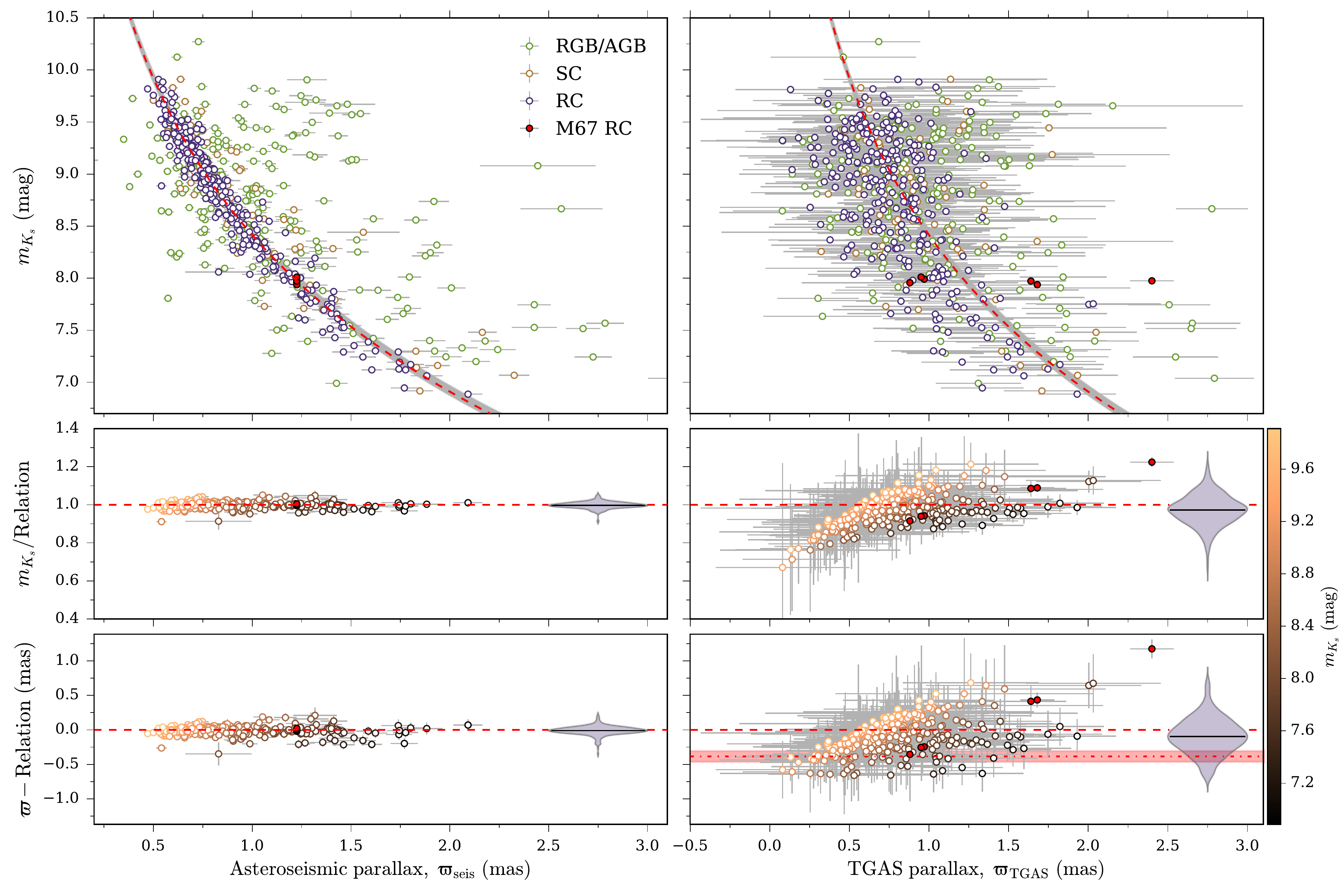}

\caption{Top: Relation between asteroseismic (left) or TGAS (right)
  parallaxes and de-reddened 2MASS $K_s$ magnitudes ($m_{\rm
    K_s}$) for our sample. The markers indicate the evolutionary state
  of the star, including the red giant branch (RGB or AGB), the red
  clump (RC), and the secondary clump (SC). We note that the uncertainties on $m_{\rm
    K_s}$ are typically smaller than the marker size. The dashed red line in
  both panels shows the theoretical relation used for the 
  residuals in the bottom panels. The relation is given by 
  \eqref{eq:rel} using the median theoretical value
  of $M_{K_s}^{RC}$ from the literature \citep[see][and references therein]{RCstars};
  the grey regions around the relations indicate the span of the
  literature values. The filled red markers denote the six asteroseismic RC
  stars in the M67 cluster \citep{M67} for which TGAS parallaxes were
  available. Middle: Fractional differences between the observed and
  predicted $m_{K_s}$ magnitudes from \eqref{eq:rel} at the asteroseismic
  (left) and TGAS (right) parallaxes of the RC stars. Bottom:
  residuals between observed and predicted parallaxes from
  \eqref{eq:rel} at the asteroseismic (left) and TGAS (right) parallaxes of
  the RC stars. The red dash-dot lines and shaded confidence regions follow the
  predicted offset of $-0.39\pm 0.08$ mas of the TGAS parallaxes by
  \citet[][]{2016arXiv160905390S}, adopting the ecliptic latitude
  $\beta=55^{\circ}\pm5$ of the \Kepler field-of-view. The inserts in
  the middle and bottom panels are violin plots of the
  distributions of the fractional differences and residuals, with the
  horizontal black lines giving the median values; the colors indicate
  the $m_{K_s}$ magnitudes. }
\label{fig:kmag}
\end{figure*}

\fref{fig:kmag} shows the results of using a sample of RC stars as
tests of the asteroseismic and astrometric distance scales.  We
calculate the theoretical dependence of the apparent $K_s$-band magnitudes
of the RC stars on distance (and hence parallax, \ie, $\varpi = 1000/d$, with $d$ in parsecs \guy{- in \fref{fig:kmag} the dashed red line}) using:
\begin{equation}\label{eq:rel}
m_{K_s}^{RC}  = \mu_0 + M_{K_s}^{RC} + A_{K_s} \, ,
\end{equation}
where $m_{K_s}^{RC}$ are apparent $K_s$ magnitudes obtained from
the Two Micron All Sky Survey
\citep[2MASS;][]{2003yCat.2246....0C,2006AJ....131.1163S};
$\mu_0=5\log_{10}(d)-5$ is the distance modulus; $A_{K_s}$ is the interstellar extinction
in the $K_s$ band, determined from $E(B-V)$ reddening values from
the 3D dust map by \citet[][]{2015ApJ...810...25G} (derived from stars
in the Pan-STARRS 1 survey) at the asteroseismic distance of a given
star and we adopted an extinction-to-reddening ratio of $R_{K_s} = 0.355\pm0.1$
\citep[][]{1999PASP..111...63F}. The maximum value of $A_{K_s}$ from the \citet[][]{2015ApJ...810...25G} map of our RC sample is $0.08$, with a mean of $A_{K_s}\approx 0.02$. $M_{K_s}^{RC}$ is the absolute
$K_s$ band magnitude for the RC population \citep[see][and
references therein]{RCstars}; we adopted a median of the literature values presented in 
\citet[][]{RCstars}, i.e., $M_{K_s}^{RC}=-1.585\pm 0.043$, 
where the uncertainty includes the RMS scatter between the different values.  
For the 6 RC stars added from the M67 cluster we de-reddened the $K_s$ 
magnitudes using $E(B-V)=0.041\pm 0.004$ \citep[][]{2007AJ....133..370T}, 
and for the asteroseismic distance we adopted the value given by 
\citet{M67} of $d=816\pm 11$ pc.  

The agreement in the left-hand panels of \fref{fig:kmag} of the
asteroseismic parallax scale with the expected relation 
demonstrates that a sample of RC stars together with
$m_{K_{s}}$ can be used to test the distance scale. There is in
contrast a clear systematic error in the TGAS parallaxes versus 
distance, even given the large scatter.

For the M67 RC stars we see noticeable differences in parallaxes 
when compared with the expected values.
We note that while it is possible to suggest a correction to the
TGAS parallaxes (see section \ref{sec:cor}), it is difficult to see how a correction based
solely on parallax or some measure of color could solve the 
problem presented by the M67 RC stars.

\begin{figure}[!ht]
\includegraphics[width=\columnwidth]{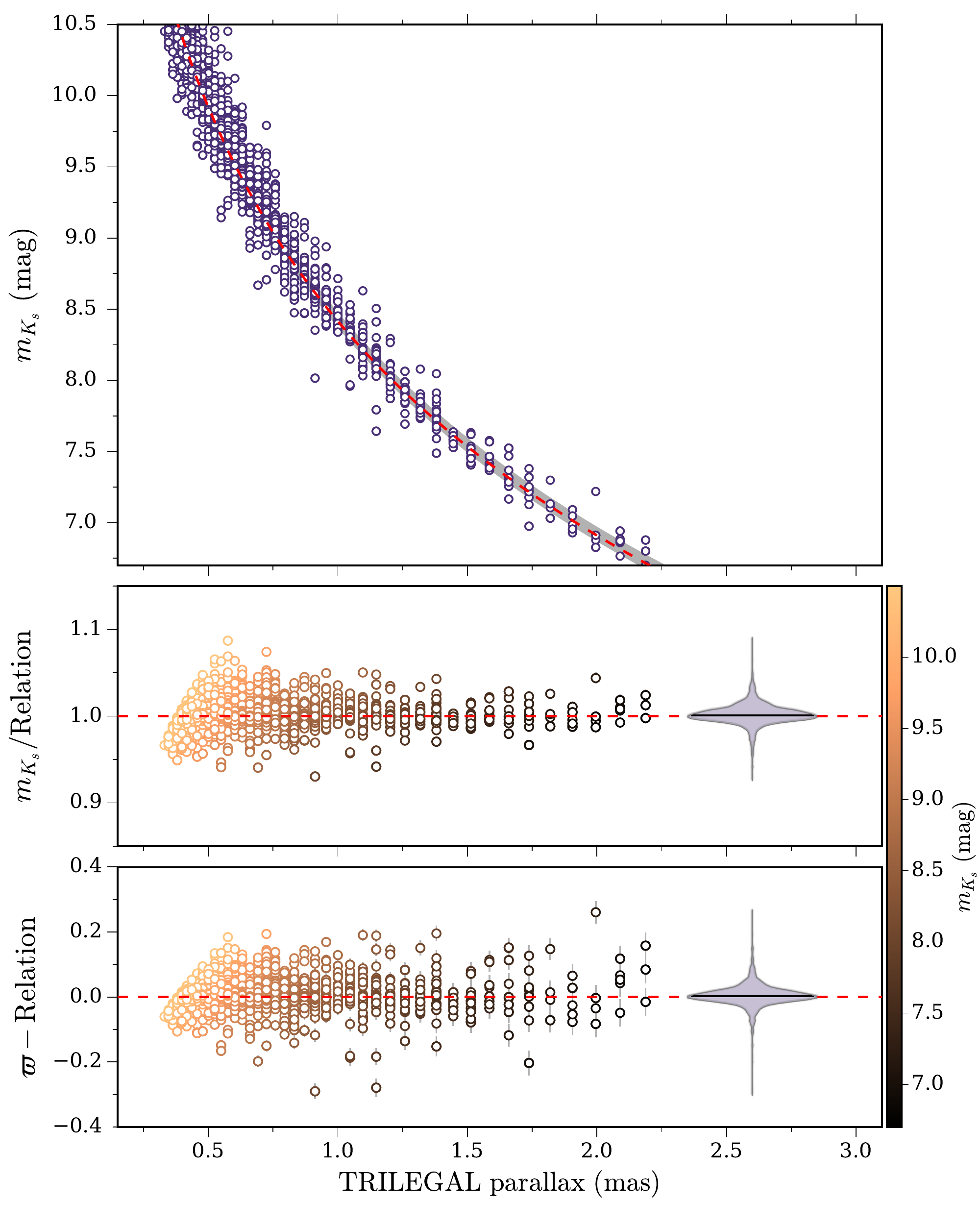}
\caption{as \fref{fig:kmag} but with a simulated RC population.}
\label{fig:Trilegal_kmag}
\end{figure}

To check for the effects of extinction, binarity, and RC
population heterogeneity on the results we have generated a
synthetic population with the properties expected of the \Kepler
field \citep[see][]{2014ApJ...784L...3M}. \fref{fig:Trilegal_kmag} plots $K_{s}$-band apparent
magnitudes as a function of parallax for a TRILEGAL simulated
population \citep[][]{2005A&A...436..895G} of single RC stars. 
The expected dependence is clear, as is the intrinsic spread in
the RC $K_{s}$ luminosity (see \eqref{eq:rel}).  The extinction
in the $K_s$ band, $A_{K_s}$, was determined from the
$A_{V}$ values in TRILEGAL and $R_{K_s}=0.355\pm0.1$ (as per
the real data) and $R_V=3.1\pm 0.1$ \citep[][]{1989ApJ...345..245C}.

As expected, for the small values of extinction expected here, 
 band extinction has
little effect on the inferred asteroseismic distances 
(when using the $K_{s}$). 
We have investigated the impact of a poorly
estimated extinction by changing the true extinction by a factor of
two.  The results of this test show that even with this large systematic
error in extinction, the impact on the estimated distances is of the
order of only a few percent. 

We have also used the synthetic
population to estimate the impact of unresolved binaries on the
apparent $K_{s}$-band magnitudes.  We find that only for a few
systems ($< 1\%$), with luminosity ratios close to unity, does the
change in apparent magnitude significantly impact the asteroseismic
distance estimate.  It is worth noting that later releases of \Gaia 
parallaxes are expected to achieve sufficiently high precision that 
the tests here could be used to identify unresolved binaries and 
constrain interstellar reddening.
\section{Suggested correction}
\label{sec:cor}
Using the sample of RC stars we propose a correction to the TGAS parallaxes. 
We find that the median offset 
between the TGAS parallaxes and the values predicted from \eqref{eq:rel}
is very close to $-0.1 \, \rm mas$ (in the sense that TGAS
overestimates the distance) which is less than either the $-0.39 \rm
\, mas$ correction for the \Kepler field ecliptic latitude or the bulk
offset of $-0.25 \rm \, mas$ given by \citet[][]{2016arXiv160905390S}.
However, we see the bias decrease at smaller parallaxes, and at parallaxes 
larger than ${\sim} 1.6 \, \rm mas$ our correction is comparable to that of \citet[][]{2016arXiv160905390S}.
\par
\begin{figure}[!h]
\includegraphics[width=\columnwidth]{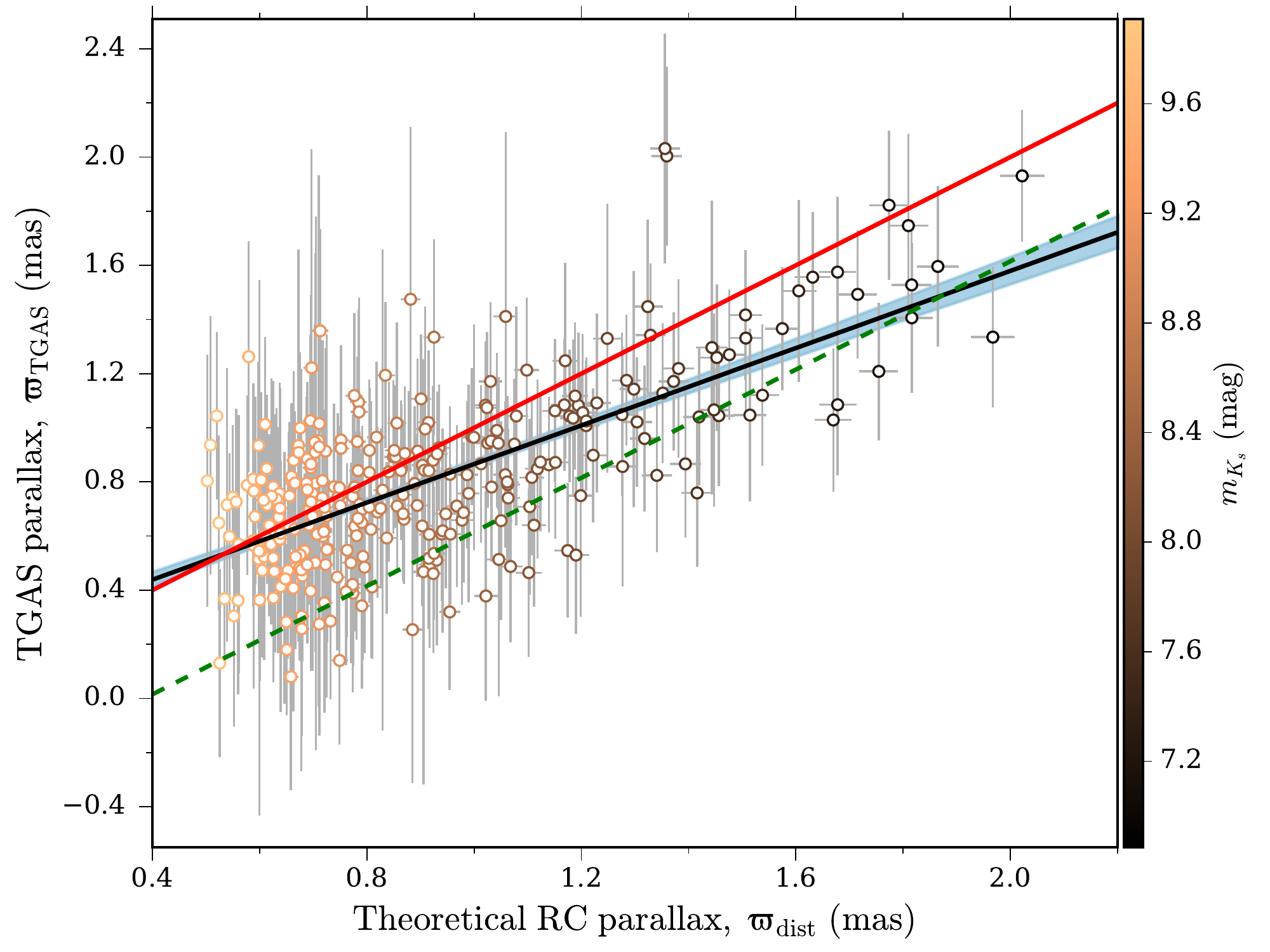}
\caption{Parallaxes predicted from \eqref{eq:rel} at the $m_{K_s}$ magnitudes of the RC stars and those of TGAS. The full red line shows the 1:1 relation. The black lines give the linear fit between the parallaxes with the associated 1-$\sigma$ uncertainty given by the blue shaded regions. The green dashed line shows the predicted off-set of $-0.39$ mas from \citet[][]{2016arXiv160905390S}, adopting an ecliptic latitude of $\beta=55^{\circ}$ of the \Kepler field-of-view. The marker colors indicate the $m_{K_s}$ magnitudes of the stars. The correction of \eqref{eq:corr} is given by the difference between the $1:1$ relation and the linear fit.}
\label{fig:kmag2}
\end{figure}

 
\fref{fig:kmag2} shows the correction we propose that
should be added to the TGAS parallaxes in the \Kepler field as a function of 
the \guy{theorectical RC parallax i.e., the parallax} that would be returned by an accurate estimate of the distance, 
$\varpi_{\rm dist}$:
\begin{equation}\label{eq:corr}\displaystyle
\Delta\varpi = \alpha +  \gamma \times \varpi_{\rm dist} \, ,
\end{equation}
with parameters $\alpha=-0.15\pm0.06$, $\gamma=0.29\pm 0.06$, 
and a parameter correlation of $r_{\alpha,\gamma} \approx -0.95$. Uncertainties 
were estimated from the ODR.  The correction may in practice be applied in an iterative
manner.  One begins by adopting $\varpi_{\rm TGAS} = \varpi_{\rm dist}$ to
compute a correction $\Delta\varpi$ using \eqref{eq:rel}, and hence a
corrected parallax. This corrected parallax is then used as an input to
\eqref{eq:rel} to compute a new correction, and an iterated corrected
parallax. The process is repeated until good convergence is found
(typically this requires only one or two iterations).

This correction is calibrated with \Kepler data and hence is strictly 
applicable to stars in the \Kepler field.
If, as suggested by \citet[][]{2016arXiv160905390S}, any correction of
TGAS parallaxes should be a function of ecliptic latitude, then the
correction could be extended to all stars at similar latitudes.  Care
should be taken if this correction is to be applied at different
latitudes.
\section{Conclusions}
\label{sec::conc}
In this letter we have used red clump (RC) stars as distance
estimators to test the systematic error in the \Gaia DR1 TGAS sample.  We
have reproduced the results of \citet[][]{joris} showing there is a
disagreement between the TGAS distances and asteroseismic distances.
By testing both results against the RC distance scale we have
demonstrated that the asteroseismic distances are in better
agreement with the RC than the TGAS distances.
The TGAS sample showed a median offset of around $-0.1 \rm \, mas$ 
(in the sense that TGAS overestimates the
distance).  We have also considered \guy{six} RC stars with
asteroseismic detections in the open cluster M67 but find
discrepancies in the TGAS parallaxes that would be challenging 
to explain in terms of a color or spatially motivated systemic error.  

We have proposed a correction to the TGAS parallaxes as a function of distance 
(as described by parallax).  
This correction is applicable to the \Kepler field-of-view \guy{at distances $> 500 \rm \, pc$, 
} but could be used at similar
ecliptic latitudes ($\beta = 55^{\circ} \pm 5$) following the findings 
of \citet[][]{2016arXiv160905390S}.  Care should be exercised when applying 
this correction to other latitudes should the TGAS bias be position dependent.
The correction we find converges to the \citet[][]{2016arXiv160905390S} correction at 
parallaxes larger that ${\sim} 1.6 \, \rm mas$ but is noticeably smaller at greater distances.

\guy{Our results then bridge the gap between the previous works that favor a correction but are sensitive to nearby
stars \citep[][]{2016arXiv160905390S, 2016arXiv161100656J}, and works that favor no correction but 
are sensitive to more distant stars \citep[][]{2016arXiv160904303L, 2016arXiv161107035S}.  We conclude
that a correction to the TGAS sample is required but that this correction becomes negligible at distances greater than $\sim 1.2 \, \rm kpc$.}
\begin{acknowledgements} 
We thank \guy{J. de Ridder, D. Huber, J. Heifetz and the anonymous referee} for their help and useful
comments.  This publication is a direct result of a hack session organised in the Solar and 
Stellar Physics group at the University of Birmingham. This work has made use of data from 
the European Space Agency (ESA) mission {\it Gaia}
(\url{http://www.cosmos.esa.int/gaia}), processed by the {\it Gaia}
Data Processing and Analysis Consortium (DPAC,
\url{http://www.cosmos.esa.int/web/gaia/dpac/consortium}). Funding for
the DPAC has been provided by national institutions, in particular the
institutions participating in the {\it Gaia} Multilateral Agreement.
The authors acknowledge the support of the UK Science and Technology
Facilities Council (STFC).  MNL acknowledges the support of The Danish
Council for Independent Research | Natural Science (Grant
DFF-4181-00415).  Funding for the Stellar Astrophysics Centre (SAC) is
provided by The Danish National Research Foundation (Grant DNRF106). LG and 
TSR acknowledge support from PRIN INAF 2014 (CRA 1.05.01.94.05)
\end{acknowledgements}
\bibliography{Bibfile}

\begin{thebibliography}{26}
\expandafter\ifx\csname natexlab\endcsname\relax\def\natexlab#1{#1}\fi

\bibitem[{{Bedding} {et~al.}(2011){Bedding}, {Mosser}, {Huber},
  {Montalb{\'a}n}, {Beck}, {Christensen-Dalsgaard}, {Elsworth},
  {Garc{\'{\i}}a}, {Miglio}, {Stello}, {White}, {De Ridder}, {Hekker}, {Aerts},
  {Barban}, {Belkacem}, {Broomhall}, {Brown}, {Buzasi}, {Carrier}, {Chaplin},
  {di Mauro}, {Dupret}, {Frandsen}, {Gilliland}, {Goupil}, {Jenkins},
  {Kallinger}, {Kawaler}, {Kjeldsen}, {Mathur}, {Noels}, {Silva Aguirre}, \&
  {Ventura}}]{2011Natur.471..608B}
{Bedding}, T.~R., {Mosser}, B., {Huber}, D., {et~al.} 2011, \nat, 471, 608

\bibitem[{Boggs \& Rogers(1990)}]{boggs1990orthogonal}
Boggs, P.~T. \& Rogers, J.~E. 1990, Contemporary Mathematics, 112, 183

\bibitem[{{Cannon}(1970)}]{1970MNRAS.150..111C}
{Cannon}, R.~D. 1970, \mnras, 150, 111

\bibitem[{{Cardelli} {et~al.}(1989){Cardelli}, {Clayton}, \&
  {Mathis}}]{1989ApJ...345..245C}
{Cardelli}, J.~A., {Clayton}, G.~C., \& {Mathis}, J.~S. 1989, \apj, 345, 245

\bibitem[{{Cutri} {et~al.}(2003){Cutri}, {Skrutskie}, {van Dyk}, {Beichman},
  {Carpenter}, {Chester}, {Cambresy}, {Evans}, {Fowler}, {Gizis}, {Howard},
  {Huchra}, {Jarrett}, {Kopan}, {Kirkpatrick}, {Light}, {Marsh}, {McCallon},
  {Schneider}, {Stiening}, {Sykes}, {Weinberg}, {Wheaton}, {Wheelock}, \&
  {Zacarias}}]{2003yCat.2246....0C}
{Cutri}, R.~M., {Skrutskie}, M.~F., {van Dyk}, S., {et~al.} 2003, VizieR Online
  Data Catalog, 2246

\bibitem[{{De Ridder} {et~al.}(2016){De Ridder}, {Molenberghs}, {Eyer}, \&
  {Aerts}}]{joris}
{De Ridder}, J., {Molenberghs}, G., {Eyer}, L., \& {Aerts}, C. 2016, ArXiv
  e-prints [\eprint[arXiv]{1609.08945}]

\bibitem[{{Elsworth} {et~al.}(2016){Elsworth}, {Hekker}, {Basu}, \&
  {Davies}}]{Elsworth2016}
{Elsworth}, Y., {Hekker}, S., {Basu}, S., \& {Davies}, G.-R. 2016, Submitted to
  MNRAS

\bibitem[{{Fitzpatrick}(1999)}]{1999PASP..111...63F}
{Fitzpatrick}, E.~L. 1999, \pasp, 111, 63

\bibitem[{{Gaia Collaboration} {et~al.}(2016){Gaia Collaboration}, {Brown},
  {Vallenari}, {Prusti}, {de Bruijne}, {Mignard}, {Drimmel}, \&
  {co-authors}}]{TGASP}
{Gaia Collaboration}, {Brown}, A.~G.~A., {Vallenari}, A., {et~al.} 2016, ArXiv
  e-prints [\eprint[arXiv]{1609.04172}]

\bibitem[{{Girardi}(2016)}]{RCstars}
{Girardi}, L. 2016, ARAA, 54, 95

\bibitem[{{Girardi} {et~al.}(2005){Girardi}, {Groenewegen}, {Hatziminaoglou},
  \& {da Costa}}]{2005A&A...436..895G}
{Girardi}, L., {Groenewegen}, M.~A.~T., {Hatziminaoglou}, E., \& {da Costa}, L.
  2005, \aap, 436, 895

\bibitem[{{Green} {et~al.}(2015){Green}, {Schlafly}, {Finkbeiner}, {Rix},
  {Martin}, {Burgett}, {Draper}, {Flewelling}, {Hodapp}, {Kaiser}, {Kudritzki},
  {Magnier}, {Metcalfe}, {Price}, {Tonry}, \&
  {Wainscoat}}]{2015ApJ...810...25G}
{Green}, G.~M., {Schlafly}, E.~F., {Finkbeiner}, D.~P., {et~al.} 2015, \apj,
  810, 25

\bibitem[{{Jao} {et~al.}(2016){Jao}, {Henry}, {Riedel}, {Winters}, {Slatten},
  \& {Gies}}]{2016arXiv161100656J}
{Jao}, W.-C., {Henry}, T.~J., {Riedel}, A.~R., {et~al.} 2016, ArXiv e-prints
  [\eprint[arXiv]{1611.00656}]

\bibitem[{{Lindegren} {et~al.}(2016){Lindegren}, {Lammers}, {Bastian},
  {Hern{\'a}ndez}, {Klioner}, {Hobbs}, {Bombrun}, {Michalik}, {Ramos-Lerate},
  {Butkevich}, {Comoretto}, {Joliet}, {Holl}, {Hutton}, {Parsons},
  {Steidelm{\"u}ller}, {Abbas}, {Altmann}, {Andrei}, {Anton}, {Bach},
  {Barache}, {Becciani}, {Berthier}, {Bianchi}, {Biermann}, {Bouquillon},
  {Bourda}, {Br{\"u}semeister}, {Bucciarelli}, {Busonero}, {Carlucci},
  {Casta{\~n}eda}, {Charlot}, {Clotet}, {Crosta}, {Davidson}, {de Felice},
  {Drimmel}, {Fabricius}, {Fienga}, {Figueras}, {Fraile}, {Gai}, {Garralda},
  {Geyer}, {Gonz{\'a}lez-Vidal}, {Guerra}, {Hambly}, {Hauser}, {Jordan},
  {Lattanzi}, {Lenhardt}, {Liao}, {L{\"o}ffler}, {McMillan}, {Mignard}, {Mora},
  {Morbidelli}, {Portell}, {Riva}, {Sarasso}, {Serraller}, {Siddiqui}, {Smart},
  {Spagna}, {Stampa}, {Steele}, {Taris}, {Torra}, {van Reeven}, {Vecchiato},
  {Zschocke}, {de Bruijne}, {Gracia}, {Raison}, {Lister}, {Marchant},
  {Messineo}, {Soffel}, {Osorio}, {de Torres}, \&
  {O'Mullane}}]{2016arXiv160904303L}
{Lindegren}, L., {Lammers}, U., {Bastian}, U., {et~al.} 2016, ArXiv e-prints
  [\eprint[arXiv]{1609.04303}]

\bibitem[{{Michalik} {et~al.}(2015){Michalik}, {Lindegren}, \&
  {Hobbs}}]{2015A&A...574A.115M}
{Michalik}, D., {Lindegren}, L., \& {Hobbs}, D. 2015, \aap, 574, A115

\bibitem[{{Miglio} {et~al.}(2014){Miglio}, {Chaplin}, {Farmer}, {Kolb},
  {Girardi}, {Elsworth}, {Appourchaux}, \& {Handberg}}]{2014ApJ...784L...3M}
{Miglio}, A., {Chaplin}, W.~J., {Farmer}, R., {et~al.} 2014, \apjl, 784, L3

\bibitem[{{Miglio} {et~al.}(2013){Miglio}, {Chiappini}, {Morel}, {Barbieri},
  {Chaplin}, {Girardi}, {Montalb{\'a}n}, {Valentini}, {Mosser}, {Baudin},
  {Casagrande}, {Fossati}, {Silva Aguirre}, \& {Baglin}}]{2013MNRAS.429..423M}
{Miglio}, A., {Chiappini}, C., {Morel}, T., {et~al.} 2013, \mnras, 429, 423

\bibitem[{{Paczy{\'n}ski} \& {Stanek}(1998)}]{1998ApJ...494L.219P}
{Paczy{\'n}ski}, B. \& {Stanek}, K.~Z. 1998, \apjl, 494, L219

\bibitem[{{Pinsonneault} {et~al.}(2014){Pinsonneault}, {Elsworth}, {Epstein},
  {Hekker}, {M{\'e}sz{\'a}ros}, {Chaplin}, {Johnson}, {Garc{\'{\i}}a},
  {Holtzman}, {Mathur}, {Garc{\'{\i}}a P{\'e}rez}, {Silva Aguirre}, {Girardi},
  {Basu}, {Shetrone}, {Stello}, {Allende Prieto}, {An}, {Beck}, {Beers},
  {Bizyaev}, {Bloemen}, {Bovy}, {Cunha}, {De Ridder}, {Frinchaboy},
  {Garc{\'{\i}}a-Hern{\'a}ndez}, {Gilliland}, {Harding}, {Hearty}, {Huber},
  {Ivans}, {Kallinger}, {Majewski}, {Metcalfe}, {Miglio}, {Mosser}, {Muna},
  {Nidever}, {Schneider}, {Serenelli}, {Smith}, {Tayar}, {Zamora}, \&
  {Zasowski}}]{2014ApJS..215...19P}
{Pinsonneault}, M.~H., {Elsworth}, Y., {Epstein}, C., {et~al.} 2014, \apjs,
  215, 19

\bibitem[{{Rodrigues} {et~al.}(2014){Rodrigues}, {Girardi}, {Miglio},
  {Bossini}, {Bovy}, {Epstein}, {Pinsonneault}, {Stello}, {Zasowski}, {Allende
  Prieto}, {Chaplin}, {Hekker}, {Johnson}, {M{\'e}sz{\'a}ros}, {Mosser},
  {Anders}, {Basu}, {Beers}, {Chiappini}, {da Costa}, {Elsworth},
  {Garc{\'{\i}}a}, {Garc{\'{\i}}a P{\'e}rez}, {Hearty}, {Maia}, {Majewski},
  {Mathur}, {Montalb{\'a}n}, {Nidever}, {Santiago}, {Schultheis}, {Serenelli},
  \& {Shetrone}}]{thiase}
{Rodrigues}, T.~S., {Girardi}, L., {Miglio}, A., {et~al.} 2014, \mnras, 445,
  2758

\bibitem[{{Salaris} \& {Girardi}(2002)}]{2002MNRAS.337..332S}
{Salaris}, M. \& {Girardi}, L. 2002, \mnras, 337, 332

\bibitem[{{Sesar} {et~al.}(2016){Sesar}, {Fouesneau}, {Price-Whelan},
  {Bailer-Jones}, {Gould}, \& {Rix}}]{2016arXiv161107035S}
{Sesar}, B., {Fouesneau}, M., {Price-Whelan}, A.~M., {et~al.} 2016, ArXiv
  e-prints [\eprint[arXiv]{1611.07035}]

\bibitem[{{Skrutskie} {et~al.}(2006){Skrutskie}, {Cutri}, {Stiening},
  {Weinberg}, {Schneider}, {Carpenter}, {Beichman}, {Capps}, {Chester},
  {Elias}, {Huchra}, {Liebert}, {Lonsdale}, {Monet}, {Price}, {Seitzer},
  {Jarrett}, {Kirkpatrick}, {Gizis}, {Howard}, {Evans}, {Fowler}, {Fullmer},
  {Hurt}, {Light}, {Kopan}, {Marsh}, {McCallon}, {Tam}, {Van Dyk}, \&
  {Wheelock}}]{2006AJ....131.1163S}
{Skrutskie}, M.~F., {Cutri}, R.~M., {Stiening}, R., {et~al.} 2006, \aj, 131,
  1163

\bibitem[{{Stassun} \& {Torres}(2016)}]{2016arXiv160905390S}
{Stassun}, K.~G. \& {Torres}, G. 2016, ArXiv e-prints
  [\eprint[arXiv]{1609.05390}]

\bibitem[{{Stello} {et~al.}(2016){Stello}, {Vanderburg}, {Casagrande},
  {Gilliland}, {Aguirre}, {Sandquist}, {Leiner}, {Mathieu}, \&
  {Soderblom}}]{M67}
{Stello}, D., {Vanderburg}, A., {Casagrande}, L., {et~al.} 2016, ArXiv e-prints
  [\eprint[arXiv]{1610.03060}]

\bibitem[{{Taylor}(2007)}]{2007AJ....133..370T}
{Taylor}, B.~J. 2007, \aj, 133, 370

\end{thebibliography}

\end{document}